\documentclass[pre,showpacs,twocolumn,floatfix,amsmath,amsfonts]{revtex4}

\usepackage{epsfig}
\usepackage{graphicx}
\usepackage{amsmath}
\usepackage{amssymb}
\usepackage{dcolumn, color}

\begin{document}

\title{Interaction of sine-Gordon kinks and breathers with a $\mathcal{PT}$-symmetric defect}

\author{Danial Saadatmand$^{1}$}
\email{saadatmand.d@gmail.com}

\author{Sergey V. Dmitriev$^{2,3}$}
\email{dmitriev.sergey.v@gmail.com}

\author{Denis I. Borisov$^{4,5}$}
\email{borisovdi@yandex.ru}

\author{Panayotis G. Kevrekidis$^{6}$}

\affiliation{ $^1$Department of Physics, Ferdowsi University of Mashhad,
91775-1436 Mashhad, Iran \\
$^2$Institute for Metals Superplasticity Problems RAS, Khalturin
39, 450001 Ufa, Russia
\\
$^3$Saint Petersburg State Polytechnical University,
Politekhnicheskaya 29, 195251 St. Petersburg, Russia
\\
$^4$Institute of Mathematics CC USC RAS, Chernyshevsky 112, 450008
Ufa, Russia
\\
$^5$  Bashkir State Pedagogical University, October Rev.  3a,
450000 Ufa, Russia
\\
$^6$Department of Mathematics and Statistics, University of
Massachusetts, Amherst, MA 01003 USA }

\begin{abstract}
The scattering of kinks and low-frequency breathers of the
nonlinear sine-Gordon (SG) equation on a spatially localized
$\mathcal{PT}$-symmetric perturbation (defect) with a balanced
gain and loss is investigated numerically. It is demonstrated that
if a kink passes the defect, it always restores its initial
momentum and energy and the only effect of the interaction with
the defect is a phase shift of the kink. A kink approaching the
defect from the gain side always passes, while in the opposite
case it must have sufficiently large initial momentum to pass
through the defect instead of being trapped in the loss region.
The kink phase shift and critical velocity are calculated with the
use of the collective variable method. Kink-kink (kink-antikink)
collisions at the defect are also briefly considered, showing how
their pairwise repulsive (respectively, attractive) interaction
can modify the collisional outcome of a single kink within the
pair with the defect. For the breather, the result of its
interaction with the defect strongly depends on the breather
parameters (velocity, frequency and initial phase) and on the
defect parameters. The breather can gain some energy from the
defect and as a result potentially even split into a kink-antikink
pair or it can lose a part of its energy. Interestingly, the
breather translational mode is very weakly affected by the
dissipative perturbation, so that a breather penetrates more
easily through the defect when it comes from the lossy side, than
a kink. In all studied soliton-defect interactions the energy loss
to radiation of small-amplitude extended waves is negligible.
\end{abstract}
 \maketitle

\section {Introduction}

The last 15 years have seen a significant series of developments
in quantum theory, stemming from the realization by
Bender and co-authors that  a class
of non-Hermitian Hamiltonians possess real spectra under the
parity-time ($\mathcal{PT}$) symmetry condition, where parity-time
means spatial reflection and time reversal, $x\rightarrow-x$ and
$t\rightarrow-t$ \cite{Bender1,Bender2}. This mathematical
discovery has initiated numerous studies of open systems with
balanced gain and loss even though the generality of this
construction is under discussion \cite{Critics}. Experimental
setups have been offered to create $\mathcal{PT}$-symmetric
physical systems in optics
\cite{Ruter,Guo,Regensburger,Regensburger1,Peng2014}, electronic
circuits \cite{Schindler1,Schindler2,Factor}, as well
as in mechanical
systems \cite{Bender3}.

In a number of theoretical studies it has been demonstrated that
$\mathcal{PT}$-symmetric systems often demonstrate unusual and
counterintuitive properties. These include, among others,
unconventional beam
refraction \cite{Zheng010103}, Bragg scattering
\cite{Longhi022102}, symmetry-breaking transitions \cite{Ruter}
and associated ghost states~\cite{wunner,graefe,usbender,vassos}, a
loss-induced optical transparency \cite{Guo}, conical diffraction
\cite{Ramezani013818}, a new type of Fano resonance
\cite{Miroshnichenko012123}, chaos \cite{West054102}, nonlocal
boundary effects \cite{Sukhorukov2148}, optical switches
\cite{Nazari} and diodes~\cite{jennie1,jennie2}, phase sensitivity of light dynamics
\cite{Barashenkov,Suchkov,Rysaeva}, and the possibility of linear
and nonlinear wave amplification and
filtering~\cite{Dmitriev013833,Suchkov033825,SuchkovEPL}.
Unexpected instabilities were also identified
at the level of $\mathcal{PT}$-symmetric lattices and nonlinear
modes were identified in few-site oligomers, as well as in full
lattice settings both in 1d~\cite{kaili,barflach2,pickton,dep1,dep2,kondep} and
even in 2d~\cite{uwe}. Extensions of $\mathcal{PT}$-symmetric
considerations in the setting of active media (of unequal
gain and loss) have also recently been
proposed~\cite{barflach,haitao}.

Motivated by the linear oscillator problems associated
with (linear) electrical~\cite{Schindler1,Schindler2}
and mechanical~\cite{Bender3} $\mathcal{PT}$-symmetric experiments,
Klein-Gordon field-theoretic generalizations with a
$\mathcal{PT}$-symmetric defect have been proposed and the
collective variable method has been developed to describe kink
dynamics in the system \cite{KevrekidisRevA}; see
also for a detailed discussion~\cite{Demirkaya2,Demirkaya}. It was also shown
that standing kinks in such models are stable if they are centered
at the loss side of the defect \cite{Demirkaya} and standing
breather may exist only if centered exactly at the interface
between gain and loss regions \cite{PanosNew}. A natural question
arises what happens with the moving Klein-Gordon solitary waves
when they interact with the spatially localized
$\mathcal{PT}$-symmetric defect.

Interaction of solitary waves with local inhomogeneities of media
has been attracting attention of researchers for the last two
decades. The reflection windows were observed in the kink-impurity
interactions by Fei {\em et al}. in the sine-Gordon (SG)
\cite{Fei1} and $\phi^4$ \cite{Fei2} models. Scattering of SG
breather by localized defects has been investigated in the
conservative case \cite{Piette}. It has been shown that the
breather can split into a kink and antikink pair or can be
accelerated by the defect. This is possible in conservative
systems because the translational kinetic energy of the breather
can be partly converted into its internal energy and vice versa.
Scattering of linear and nonlinear waves (solitons) on defects in
$\mathcal{PT}$-symmetric optical waveguide arrays was analyzed
\cite{Dmitriev013833,Suchkov033825,SuchkovEPL,jennie1,jennie2}. It was
shown that
the incident high-amplitude solitons (or even linear
wavepackets~\cite{jennie1,jennie2}) can excite a mode localized
on the $\mathcal{PT}$-symmetric defect. By exciting the localized
mode of a large amplitude, it is possible to perform
phase-sensitive control of soliton scattering and amplification or
damping of the localized mode. The gain-loss pattern
in conjunction with the nonlinearity lead to asymmetric propagation
of the incoming wavepackets depending on their direction of
incidence.

Kinks in non-integrable models such as the perturbed SG equation
or $\phi^4$ model can support internal vibrational modes
\cite{KivsharIM}. In some cases, impurities can also support
localized vibrational modes. Kinks of the integrable SG equation
do not bear internal modes \cite{Quintero}. When a kink hits an
impurity in a conservative model, a part of its energy is trapped
towards the excitation of the impurity mode
\cite{Fei1,Kivshar1177} and another fraction leads to the emission
of radiation bursts \cite{Malomed385}.

A merger of a colliding kink and antikink into a breather is
possible in a non-integrable system when energy loss to radiation
and/or excitation of the kink's internal modes is sufficiently
large \cite{Campbell1,Goodman,Campbell2,Campbell3}. The binding
free kink and antikink into a breather has been addressed in
\cite{Scharf} in presence of spatially periodic perturbation.
External d.c. driving force in the absence of damping for
sufficiently large magnitude of the force causes the breather to
split into a kink-antikink pair while for small driving force the
breather excitations lead to stationary modes \cite{Lomdahl}. The
breather can dissociate into a kink-antikink pair under external
field \cite{Inoue}. Conversion of an oscillation mode into a
kink-antikink pair has been observed via abrupt distortions of the
on-site potential in time or in space \cite{Carvalho}. The recent
work of~\cite{PanosNew} illustrated that such an evolution
is also possible when the breather is subject to gain
e.g. on the gain side of a $\mathcal{PT}$-symmetric medium.

Interaction of moving solitons with $\mathcal{PT}$-symmetric
defects in the realm of the Klein-Gordon field, to the best of our
knowledge, has not been studied previously,
in part, arguably, since  $\mathcal{PT}$-symmetric
field theories is a very recent theme of research.
In this paper, we aim to reveal
the principal physical effects observed during the interaction of SG
kinks and breathers with a $\mathcal{PT}$-symmetric defect with
balanced gain and loss. In particular, a kink approaching the
$\mathcal{PT}$-symmetric defect from the gain side is always transmitted,
while from the loss side it may be reflected or transmitted
depending on its energy. This suggests an asymmetric effective
dynamics which is identified by means of an explicitly solvable
collective coordinate approach. We also illustrate how this effective
collective dynamics can be modified by the presence of the repulsion
from an another kink or of the attraction by an antikink.
On the other hand, for the breather the dynamics is sensitively
dependent on both the characteristics of the breather and those
of the defect, potentially exhibiting either gain or loss of
energy for the coherent structure (the former possibly even featuring
the breakup of the breather into a kink and an anti-kink waveforms).

The structure of the paper is as follows. In Sec.~\ref{Sec:II},
following the work \cite{KevrekidisRevA}, we introduce the
perturbed SG equation and the well-known kink and breather
solutions to the integrable SG equation. In Sec.
\ref{Sec:CollectVar}, a collective variable method is applied
and analytically solved to
reveal some features of the kink dynamics in the considered
system. We then report on the numerical results for scattering of
kinks in Sec.~\ref{Sec:Kink}, pairs of kinks in
Sec.~\ref{Sec:KinkKink}, breathers in Sec.~\ref{Sec:Breather} and
kink-antikink pairs in Sec.~\ref{Sec:KinkAntikink}. Our Conclusions
and some future directions are presented in Sec. \ref{Sec:V}.

\section {The model} \label{Sec:II}
We consider a perturbed sine-Gordon equation of the form
\cite{KevrekidisRevA}:
\begin{equation}\label{SGE_Collisions}
\phi_{tt} - \phi_{xx} + \sin \phi = A\gamma(x)\phi_{t},
\end{equation}
where $\phi(x,t)$ is the unknown scalar field and lower indices
denote partial differentiation. The perturbation term is in the
right-hand side of the equation. The parameter $A$ controls the
perturbation amplitude. In order to study the effects of a
spatially localized $\mathcal{PT}$-symmetric defect on traveling
kinks and breathers, for the function $\gamma(x)$ we take
\begin{equation}\label{SGE_Perturbation}
\gamma(x)=\{\exp[-\beta(x+\delta)^2]-\exp[-\beta(x-\delta)^2]\},
\end{equation}
which has the symmetry $\gamma(-x)=-\gamma(x)$.
This ensures that Eq.~(\ref{SGE_Perturbation})
is $\mathcal{PT}$-symmetric physically
implying that while Eq. (\ref{SGE_Collisions}) describes an open
system with gain and loss, the gain balances the loss. The
gain-loss spatial profile determined by
Eq. (\ref{SGE_Perturbation}) represents a superposition of two
bell-shaped functions with the separation between them controlled by
the parameter
$\delta$. The parameter $\beta$ is related to the hump inverse width. For
$A=0$ and/or $\delta=0$ one has $\gamma(x)\equiv 0$. For
definiteness, here we consider the case of $\delta>0$ with the
gain (loss) region $x<0$ ($x>0$).

In the present study the simulations are carried out for different
values of the perturbation amplitude $A$ and fixed $\beta=0.5$ and
$\delta=0.1$. The choice of $\beta=0.5$ makes the hump width
comparable to the kink width.

For $\gamma(x)\equiv 0$, we have the integrable SG equation with
the following kink solution
\begin{equation}\label{Kink}
 \phi_K(x,t)=4\arctan\{\exp[\delta_k(x-x_0-V_k t)]\},
\end{equation}
and the breather solution
\begin{equation}\label{Breather}
   \phi_B(x,t)=4\arctan\frac{\eta\sin\{-\delta_b\omega[t-V_b(x-x_0)]\}}{\omega\cosh[\delta_b\eta(x-x_0-V_bt)]},
\end{equation}
where $V_k$ is kink velocity, $V_b$, $\omega$ are the breather
velocity and frequency, $x_0$ is the soliton initial position and
\begin{equation}\label{deltaeta}
   \delta_{k,b} =\frac{1}{ \sqrt{1 - V_{k,b}^2}},\quad \eta =
   \sqrt{1-\omega^2}.
\end{equation}
The energy of the kink and the breather are, respectively
\begin{equation}\label{Energy}
  E_k=8\delta_{k},\quad E_b=16\eta\delta_{b}.
\end{equation}

Far from the defect, solitons move with constant velocities $V_k$
and $V_b$ feeling no perturbation. In the vicinity of the defect, the
soliton parameters change and, as it will be shown, it is
important from which side the soliton hits the defect.

To study numerically the effect of the perturbation on the
dynamics of the SG solitons, we introduce the mesh $x=nh$, where
$h$ is the lattice spacing, $n=0,\pm 1,\pm 2...$ and propose the
following discrete version of the model
\begin{eqnarray}\label{FK_Collisions}
  && \frac{d^2\phi_n}{dt^2} - \frac{1}{h^2}
   \left(\phi_{n-1} - 2\phi_n + \phi_{n+1}\right)+ \nonumber\\
   &&\frac{1}{12h^2} \left(\phi_{n-2} -4\phi_{n-1}+6\phi_n -4\phi_{n+1}+ \phi_{n+2}\right) \nonumber\\
   &&+ \sin \phi_n -A\gamma_n\frac{d\phi_n}{dt}= 0,
\end{eqnarray}
in which $\phi_n=\phi(nh,t)$ and $\gamma_n=\gamma(nh)$. It can be
seen that the term $\phi_{xx}$ in Eq. (\ref{SGE_Collisions}) is
discretized with the accuracy $O(h^4)$ which has already been used
by other authors \cite{BraunKivshar,KivsharMalomed}. This is done
to minimize the effect of discreteness introduced by the mesh.
Equations of motion (\ref{FK_Collisions}) were integrated with
respect to the temporal variable using an explicit scheme with the
accuracy of $O(\tau^4)$ and the time step $\tau$. The simulations
reported below in Section IV were conducted for $h=0.1$ and $\tau=0.005$.

\section {Collective variable method} \label{Sec:CollectVar}

A collective variable approach has been developed
\cite{KevrekidisRevA} to describe the kink dynamics in the model
Eq. (\ref{SGE_Collisions}). The kink is effectively described by
the one degree of freedom particle of mass $M=8$, which is the
mass of standing kink. The kink coordinate $X(t)=x_0-V_k t$ as a
function of time $t$ can be found from the following equation of
motion
\begin{eqnarray}\label{Coll_var1}
    M\ddot{X}=A\dot{X}f(X),
\end{eqnarray}
with
\begin{eqnarray}\label{Coll_var1a}
    f(X)&=&\int_{-\infty}^{\infty} [\phi^{\prime}_K
    (x-X)]^2\gamma(x)dx,
\end{eqnarray}
where the overdot means differentiation with respect to time and
the prime denotes differentiation with respect to $X$.
Substituting the kink solution Eq. (\ref{Kink}) into Eq.
(\ref{Coll_var1a}) one obtains
\begin{eqnarray}\label{Coll_var2}
    f(X)&=&4\delta_k^2\int_{-\infty}^{\infty}\frac{\gamma(x)dx}{{\cosh}^2[\delta_k(x-X)]}.
\end{eqnarray}

The equation of motion (\ref{Coll_var1}) was integrated
numerically for the initial conditions $X(0)=x_0$,
$\dot{X}(0)=V_k$ using the simplest scheme
\begin{eqnarray}\label{Coll_var3}
    X_{i+1}=\frac{2X_i-(1+a_i)X_{i-1}}{1-a_i},
\end{eqnarray}
where $i$ denotes the time step number, $a_i=Af_i\tau/(2M)$,
$f_i=f(X_i)$, and $\tau=0.005$ is the time step.

The collective variable equation (\ref{Coll_var1}) can also be solved
explicitly with its solution given in the form of
a quadrature. The first integral reads
\begin{eqnarray}\label{Denis3}
 M\dot{X}=AF(X)+C_1,
\end{eqnarray}
where
\begin{eqnarray}\label{Denis4}
 F(X)=\int_0^X f(t)dt,
\end{eqnarray}
and $C_1$ is the integration constant. The second integration
gives
\begin{eqnarray}\label{Denis6}
\int_0^X \frac{Mdz}{AF(z)+C_1}=t+C_2,
\end{eqnarray}
with the integration constant $C_2$. Equation (\ref{Denis6}) gives
the solution to Eq. (\ref{Coll_var1}) in an implicit form
$t=t(X)$.

\subsection {Kink's phase shift due to interaction with the defect} \label{Sec:CollectVarPS}

The kink approaching the defect from the gain (loss) side is first
accelerated (decelerated) and then decelerated (accelerated) when
it enters the lossy (gain) side. As a result, the kink experiences
a phase shift. To calculate the phase shift we substitute Eq.
(\ref{Coll_var2}) into Eq. (\ref{Denis4}):
\begin{eqnarray}\label{Denis7}
F(X)=\int_0^Xds\int_{-\infty}^{\infty}
\frac{\Gamma(x)dx}{\cosh^2[\delta_k(x-s)]},
\end{eqnarray}
where $\Gamma(x)=4\delta_k^2\gamma(x)$.
The function
\begin{eqnarray}\label{Denis8}
f(s)=\int_{-\infty}^{\infty}\frac{\Gamma(x)dx}{\cosh^2[\delta_k(x-s)]}
\end{eqnarray}
is odd and hence the function $F(X)$ is even. Note that $f(s)$
decays exponentially when $s\rightarrow \pm \infty$. From the last
statement it follows the existence and the equality of the
following limits
\begin{eqnarray}\label{Denis9}
B= \lim_{s\to+\infty}F(s)=\lim_{s\to-\infty}F(s),
\end{eqnarray}
where $B$ is the value of the limits.

Coming back to Eq. (\ref{Denis6}), we note that the integrand can
be presented as the sum
\begin{equation}\label{Denis12}
\frac{M}{AF(z)+C_1}=\frac{M}{AB+C_1}+\frac{MA(B-F(z))}{(AF(z)+C_1)(AB+C_1)}.
\end{equation}
Substitution of the last equation into Eq. (\ref{Denis6}) gives
\begin{equation}\label{Denis13}
\frac{M X(t)}{AB+C_1}+\int_0^X
\frac{MA(B-F(z))}{(AF(z)+C_1)(AB+C_1)}dz=t+C_2.
\end{equation}
The integral in Eq. (\ref{Denis13}) is bounded uniformly in $X$
since the integrand decays exponentially at infinity. The right
hand side in (\ref{Denis13}) is a linear function in $t$. Hence,
as $t\to\pm\infty$, each solution to equation (\ref{Coll_var1})
should behave as
\begin{equation}\label{Denis21}
X(t)=V_kt+O(1),
\end{equation}
where 
$O(1)$ indicates terms bounded as $t\to\pm\infty$, and
$V_k$ is in fact the kink velocity given by the formula
\begin{equation}\label{Denis11}
\frac{1}{V_k}=\frac{M}{AB+C_1}.
\end{equation}
The last equation expresses
the kink velocity in terms of the model parameters. Below we
assume that $V_k>0$, and the case of $V_k<0$ can be treated in a
similar way.

The function $X(t)$ Eq.~(\ref{Denis21}) grows at infinity linearly and
hence the integral in the left hand side in (\ref{Denis13}) tends
to a constant as $t\to\pm\infty$. Thus, we can specify behavior of
Eq.~(\ref{Denis21}) as follows,
\begin{eqnarray}\label{Denis14}
X(t)=V_kt+x_\pm+o(1),\quad t\to\pm\infty,
\end{eqnarray}
where now the symbol $o(1)$  stands for the terms vanishing as
$t\to\pm\infty$.

The quantity $\Delta x=x_+-x_-$ is in fact the kink's phase shift
due to the defect, which we now calculate. In order to do it, we
substitute Eq. (\ref{Denis14}) into Eq. (\ref{Denis13}), taking
into consideration Eq. (\ref{Denis11}):
\begin{eqnarray}\label{Denis15}
\frac{x_\pm}{V_k}+o(1)+\frac{1}{V_k}\int_0^{X(t)}
\frac{A(B-F(z))}{AF(z)+C_1}dz=C_2.
\end{eqnarray}
In the limit $t\to\pm\infty$, for positive $V_k$ one has $X(t)\to
\pm\infty$, and Eq. (\ref{Denis15}) becomes
\begin{eqnarray}\label{Denis16}
\frac{x_\pm}{V_k}+\frac{1}{V_k}\int_0^{\pm\infty}
\frac{A(B-F(z))}{AF(z)+C_1}dz=C_2.
\end{eqnarray}
Subtracting one identity from the other one, finds
\begin{eqnarray}\label{Denis17}
\Delta x=x_+-x_-=-\int_{-\infty}^{+\infty}
\frac{A(B-F(z))}{AF(z)+C_1}dz.
\end{eqnarray}
The integration constant $C_1$ can be found from Eq.
(\ref{Denis11}) that allows us to rewrite Eq. (\ref{Denis17}) as
\begin{equation}\label{Denis19}
\Delta x=\int_{-\infty}^{+\infty}
\frac{A(F(z)-B)}{A(F(z)-B)+MV_k}dz.
\end{equation}
If $V_k<0$, the similar formula reads as
\begin{equation}\label{Denis22}
\Delta x=-\int_{-\infty}^{+\infty}
\frac{A(F(z)-B)}{A(F(z)-B)+MV_k}dz.
\end{equation}
For the kink solution (\ref{Kink}) the function $F(z)-B$ can be cast into the particular form
\begin{equation}\label{Denis20}
F(z)-B=-\int_z^{+\infty} ds\int\limits_{-\infty}^{+\infty}
\frac{\Gamma(x)dx}{\cosh^2[\delta_k(x-s)]}.
\end{equation}
After changing the order of integration and integrating over $s$
one obtains
\begin{equation}\label{Denis23}
F(z)-B=-\delta_{k}^{-1}\int\limits_{-\infty}^{+\infty}
\Gamma(x)\{1-\tanh[\delta_k(z-x)]\}dx.
\end{equation}
The kink's phase shift can be now found from Eqs. (\ref{Denis19}), (\ref{Denis22})
and Eq. (\ref{Denis23}) by evaluating the integrals numerically.

\subsection {Critical kink velocity} \label{Sec:CollectVarKV}

If the kink approaches the defect from the loss side, it must have
sufficient momentum not to be trapped. The critical kink initial
velocity $V_c$ can be found with the help of the collective
variable method. One can present Eq.~(\ref{Coll_var1}) for
$\dot{X}$ in the form
\begin{eqnarray}\label{Coll_var4}
    M(\dot{X}-\dot{X_0})=\delta_k A\int_{-\infty}^{\infty}\int_{X_0}^{X} \frac{\Gamma(x)dxdX} {{
    \cosh}^2[\delta_k(x-X)]}.  \nonumber \\
\end{eqnarray}
A kink having critical velocity must have $\dot{X}=0$ at $X=0$,
i.e., the kink stops when it reaches the center of the defect.
Setting in Eq. (\ref{Coll_var4}) $V_c=\dot{X_0}$ and $\dot{X}=0$
after integrating over the the collective variable $X$ we have
\begin{eqnarray}\label{Coll_var5}
    V_c=\frac{A}{\delta_k M}\int_{-\infty}^{\infty}\Gamma(x) \{{\tanh}[\delta_k(x-X_0)]-{
    \tanh}[\delta_k(x-X)]\}dx. \nonumber \\
\end{eqnarray}
The value of the integral in Eq. (\ref{Coll_var5}) can be found
numerically for the initial condition $X_0=15$ and recalling that
the final stopping point is ${X}=0$. For
$\beta=0.5$ and $\delta=0.1$ used in our study one finds
$V_c=0.3066(4A/\delta_k M)$. For small kink velocity $\delta_k=1$
and $M=8$ so that
\begin{eqnarray}\label{Vc}
   V_c=0.1533A.
\end{eqnarray}

\section {Numerical Results} \label{Sec:IIA}

\subsection {Kink-defect interaction} \label{Sec:Kink}

First we start with the case of the kink-defect interaction, which
is simpler.

\begin{figure}
\includegraphics[width=9.5cm]{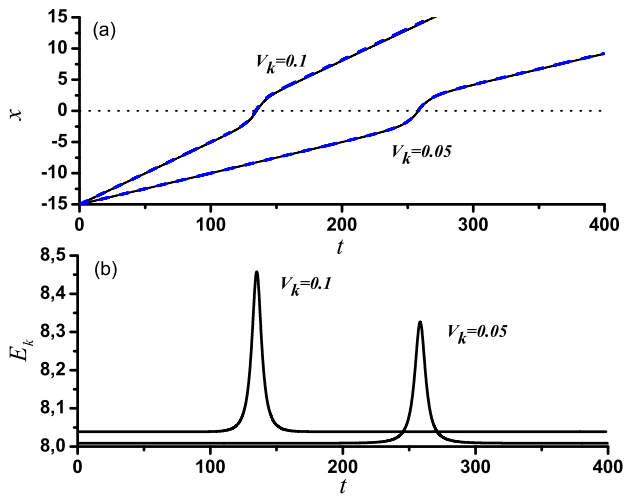}
\caption{(a) Kink position as a function of time for the two
values of kink initial velocity $V_k=0.05$ and $V_k=0.1$, for the
case when the kink approaches the defect from the gain side.
The defect center is located at $x=0$. Solid lines show the results of
the numerical solution for the continuous system and dashed lines show
the results obtained with the help of the collective variable
method. Horizontal dotted line shows location of the defect
center. (b) Time evolution of the kink total energy with the
initial velocities $V_k=0.05$ and $V_k=0.1$ during the interaction
with the defect. The perturbation amplitude is $A=1.5$ in both
cases.} \label{fig1}
\end{figure}

In Fig.~\ref{fig1} the results for the case when the kink
approaches the defect with $A=1.5$ from the gain side are
presented. In (a) the kink position as the function of time is
shown by the solid lines for the two values of the initial kink
velocity, $V_k=0.05$ and $V_k=0.1$, as indicated for each curve.
Dashed lines give the results obtained with the help of the theoretical
collective variable method Eq. (\ref{Coll_var3}). One can see that
the collective variable approach gives a very accurate prediction
of the actual kink dynamics. In (b) time evolution of the kink
energy $E_k$ is plotted. From Fig.~\ref{fig1} it is clearly seen
that the kink moving toward the defect from the gain side is first
accelerated and after passing the gain side of the defect it is
decelerated by the loss side. After the kink passes the defect and
moves far from it, it restores its initial velocity and energy.
The only effect of the kink-defect interaction in this case is a
phase shift. The maximal kink energy increases with increase in
the kink initial velocity $V_k$ for fixed defect amplitude $A$
because of the nature of the defect, whose effect is stronger for
larger $\phi_t$.

\begin{figure}
\includegraphics[width=9.5cm]{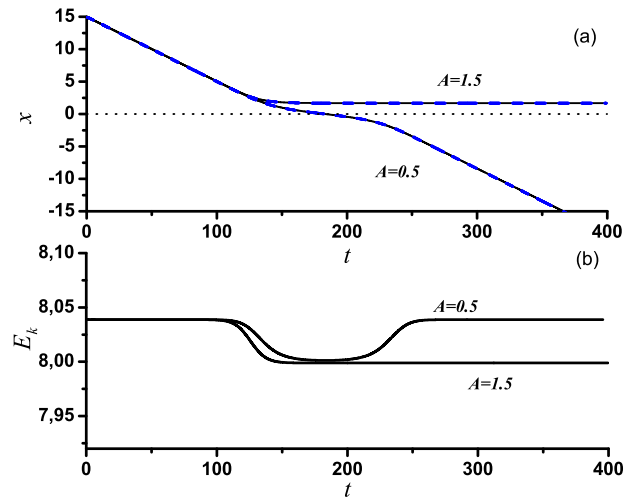}
\caption{(a) Kink position as a function of time for the case when
the kink approaches the defect from the loss side. The kink velocity
is $V_k=-0.1$ and the results are given for $A=0.5$ and $A=1.5$.
The results for the full system described by the
partial differential equation (PDE) of Eq.~(\ref{SGE_Collisions})
(solid lines) and the ordinary differential equation (ODE) of the collective
variable approach (dashed lines) are compared. The dotted line shows the
location of the defect center. (b) Time evolution of the kink
total energy for the same two cases.} \label{fig2}
\end{figure}

Next, suppose a kink comes from the lossy side. In this case, two
different scenarios for the kink interaction with the defect are
possible depending on its initial velocity $V_k$ (or perturbation
strength $A$) as presented in Fig.~\ref{fig2}. If $V_k$ is large
enough (or $A$ is small enough), the kink passes through the lossy
part of the defect with the velocity smaller than $V_k$ and enters
the gain part where it is accelerated up to the initial velocity
and then goes on to infinity. In the opposite case (where $V_k$ is not
large enough or $A$ is not small enough), the kink does not possess
sufficient momentum to pass through the lossy part of the defect
and it is trapped there. In Fig.~\ref{fig2}~(a) the kink position as
a function of time is shown for $A=0.5$ and $A=1.5$ with
$V_k=-0.1$ in both cases. Figure~\ref{fig2}~(b) shows the kink
total energy as a function of time for these two cases. As one
can see, for the case of $A=0.5$ kink passes through the defect
and restores its initial velocity, while for $A=1.5$ the kink is
trapped by the lossy side of the defect.
\begin{figure}
\includegraphics[width=9.5cm]{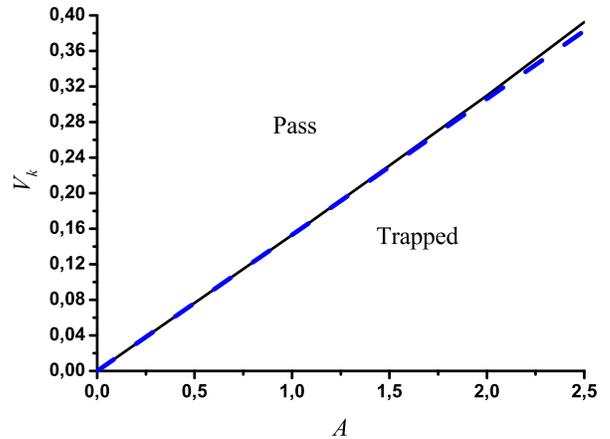}
\caption{Relation between the critical initial velocity of the
kink and the critical defect amplitude. Below the line
$V_k=0.153A$ the kink approaching the defect from the lossy side
is always trapped there. Above the line it passes through the
defect and restores its initial velocity. Solid line is for the
continuum system, while dashed line is for collective variable
method.} \label{fig3}
\end{figure}

We now further expand on our comparison of the kink dynamics observed in the
continuous PDE system of Eq. (\ref{SGE_Collisions}) with that in the
single degree of freedom ODE model of Eq. (\ref{Coll_var1}).
In Fig. \ref{fig3} the plane of the parameters $A$ and $V_k$ is
shown with the line which separates the two possible scenarios of
the kink-defect interaction when the kink approaches the defect
from the lossy side. Above the line the kink has sufficient
initial momentum to pass through the defect and to restore its
initial velocity. Contrary to this, below the line the kink is
always trapped in the lossy region of the defect and eventually
stops. The collective variable result Eq.~(\ref{Vc}) is shown by
the dashed line. The result obtained for the continuum system
(shown by the solid line) is in a perfect agreement with the
collective variable method for small kink velocity and the
deviation increases for larger kink velocities. This is
natural to expect as the collective coordinate derivation
of~\cite{KevrekidisRevA} was obtained away from the relativistic
regime of large speeds $V_k$. Nevertheless, we observe that for
speeds even nearly half the maximal speed of propagation in the
medium the relevant collective coordinate prediction remains
very accurate.

\begin{figure}
\includegraphics[width=9.0cm]{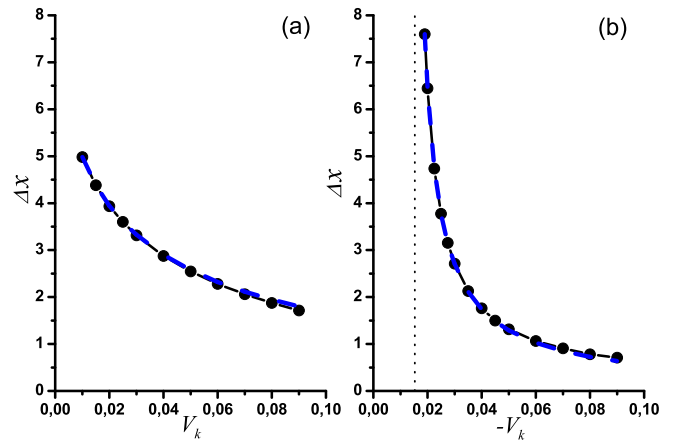}
\caption{Kink's phase shift due to the interaction with a
$\mathcal{PT}$-symmetric defect as a function of its initial
velocity for the kink moving (a) toward the gain side of the
defect with strength $A=0.5$ and (b) toward the lossy side of the
defect with strength $A=0.1$. Solid lines show the results of
numerical solution for the continuous PDE system and dashed lines show
the results obtained with the help of the collective variable ODE
method Eqs. (\ref{Denis19}), (\ref{Denis22}) and (\ref{Denis23}).
The vertical dotted line in (b) shows the threshold kink velocity
$V_c=-0.0153$.} \label{fig4}
\end{figure}

In Fig.~\ref{fig4}, the kink's phase shift due to interaction with the
$\mathcal{PT}$-symmetric defect is presented as a function of its
initial velocity. Solid lines show the results of the numerical
solution for the continuous system and dashed lines show the results
obtained with the help of the collective variable method of Eqs.
(\ref{Denis19}), (\ref{Denis22}), and (\ref{Denis23}). In (a) the
kink moves toward the defect with strength $A=0.5$ from the gain
side. In (b) the kink moves from the opposite side and $A=0.1$.
The vertical dotted line shows the critical value of the initial
kink velocity for this case. It can be seen that the accuracy of
the collective variable method is very high especially for small
$V_k$. The plots show smaller phase shift for higher kink initial
velocity. This comes from the fact that higher velocity kink is
more accelerated by the perturbation considered here.

\subsection {Kink-kink-defect interaction} \label{Sec:KinkKink}

\begin{figure}
\includegraphics[width=9cm]{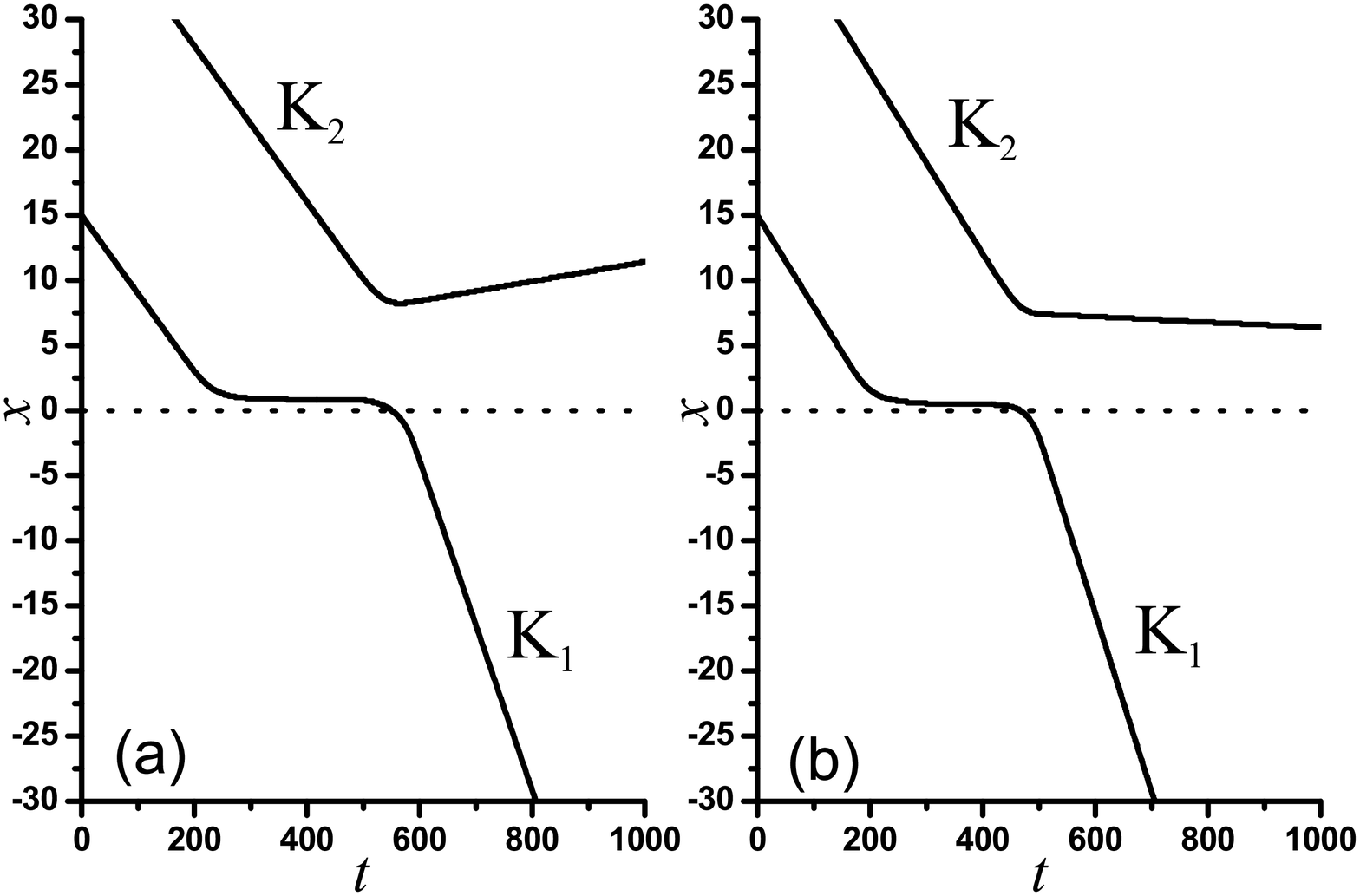}
\caption{Dynamics of the two well separated kinks moving toward
the defect from the lossy side with a velocity smaller than the
threshold value. K$_1$ is trapped by the defect and then it is
pushed through the defect by K$_2$, through their mutual
repulsion. In (a) K$_2$ is reflected,
while in (b) it is trapped by the lossy region of the defect.
Horizontal dashed lines show the location of the center of the
defect. Here (a) $V_k=-0.06$, (b) $V_k=-0.07$ and $A=0.5$ in both
cases. The threshold kink velocity is $V_c=-0.0765$. The kink initial
positions are $x_0=15$ for K$_1$ and $x_0=40$ for K$_2$.}
\label{fig5}
\end{figure}

Here we demonstrate that the kink K$_1$ trapped at the lossy side
of the defect can be pushed through the defect by the second kink
K$_2$ even if the second kink has velocity smaller than the
threshold value. To do so we consider two well separated kinks
moving with the same velocity below the threshold value toward the
lossy side of the defect. The first kink is trapped and the second
one pushes it, through their well-known mutual repulsion~\cite{KivsharMalomed},
through the defect being either reflected back [see
Fig.~\ref{fig5}~(a) for the case of $V_k=-0.06$] or trapped itself [as in
Fig.~\ref{fig5}~(b) for the case of $V_k=-0.07$]. Note that the
threshold kink velocity is $V_c=-0.0765$ for $A=0.5$, used for
this simulation.

\subsection {Breather-defect interaction} \label{Sec:Breather}

It was found that the result of the breather-defect interaction importantly
depends on the initial breather position $x_0$, because this
parameter controls the breather oscillation phase at which it hits
the defect. A moving breather in one oscillation travels the
distance $\lambda=2\pi\delta_b|V_b|/\omega$. This means that it is
sufficient to consider the range of the initial breather positions
from $x_0$ to $x_0+\lambda$. In some cases the breather can split
into kink and antikink after passing the defect. In this situation
it is convenient to present the result of the breather-defect
interaction by the total energies of the subkinks constituting the
breather under the assumption that the subkinks, when
merged into a resulting breather after the interaction, share the breather
energy equally. In the cases when the breather splits into a
kink-antikink pair, the energies of the subkinks are different and
they are calculated after they become well separated.

\begin{figure}
\includegraphics[width=12cm]{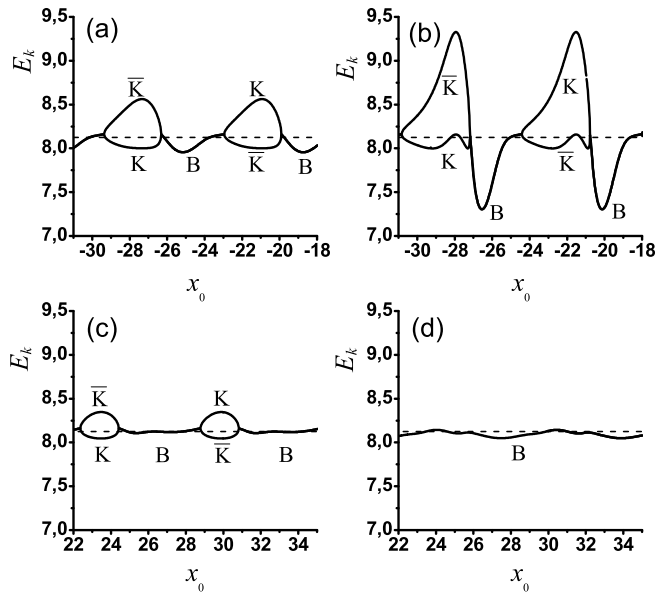}
\caption{Numerical results for the breather interaction with the
$\mathcal{PT}$-symmetric defect in the perturbed SGE model Eq.
(\ref{SGE_Collisions}) for the breather approaching (a,b) from the
gain side of the defect and (c,d) from the lossy side of the
defect. Shown are the total energies of the subkinks constituting
the breather under the assumption that they share the breather
energy equally. When the breather splits into a kink-antikink
pair, the subkinks have different energies and the lines split
into two. The defect amplitude is $A=0.1$ in (a,c) and $A=0.4$ in
(b,d). The breather has initial velocity $V_b=\pm 0.2$ and
frequency $\omega=0.1$. } \label{fig6}
\end{figure}

In Fig. \ref{fig6} we plot the total energy of the subkinks after
the breather collides with the defect as a function of its initial
position $x_0$. Horizontal dashed lines show the initial energy of
the subkinks. In (a,b) we show the case when the breather
approaches the defect from the gain side and in (c,d) it moves
toward the defect from the opposite direction. In (a,c) the defect
amplitude is A=0.1, while in (b,d) A=0.4. The breather has
frequency $\omega=0.1$ and initial velocity of $V_b=\pm 0.2$, so
that in all cases $\lambda=12.83$. The plots include the whole
period of the breather initial position. One can see that in (a-c)
there exist the domains of $x_0$ where the breather (B) splits
into a kink-antikink (${\rm K}-\overline{\rm K}$) pair. In (d) the
breather does not gain enough energy from the defect to split. It
can be concluded that the breather can split regardless of the
direction it approached the defect. However, the maximal energy
gain is larger when the breather moves toward the gain side of the
defect. Interestingly also, although there are (naturally expected)
cases where the splinters bear a lower energy sum than that of
the original breather, there are also ones where their sum
exceeds the energy of the original breather. Again, this can
happen on either side of approach, although it is again more
pronounced when approaching the defect from the gain side.

\begin{figure}
\includegraphics[width=9cm]{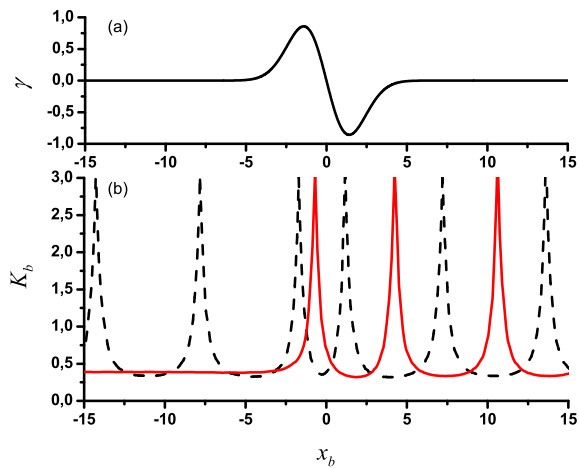}
\caption{(a) $\gamma$ as the function of $x$. (b) The kinetic
energy of the breather as a function of its position. The breather
comes from the lossy (positive $x_b$, i.e., right end) side of the defect
towards the gain (negative $x_b$, i.e., left end) side with initial positions
$x_{0}=23.5$ (solid line) and $x_{0}=26.5$ (dashed line) [see
Fig.~\ref{fig6}(c)]. In both cases $V_b=-0.2$, $\omega=0.1$ and
$A=0.1$.} \label{fig7}
\end{figure}

To explain the dependence of the breather-defect interaction on
the breather phase we note that the kinetic energy of the moving
breather is a periodic function of time and space with the sharp
maxima at the points where the subkinks collide. The perturbation
term in Eq.~(\ref{SGE_Collisions}), as it was already mentioned,
acts more prominently for large $\phi_t$. Thus, the location of
the subkink collision points with respect to the maximum and
minimum of $\gamma(x)$ is very important. In Fig.~\ref{fig7}(a)
the function $\gamma(x)$ is shown. In Fig.~\ref{fig7}(b) the
breather kinetic energy $K_b$ as the function of its spatial
coordinate $x_b$ is given for the two cases, $x_{0}=23.5$ (solid
line) and $x_{0}=26.5$ (dash line), for $V_b=-0.2$ and
$\omega=0.1$, which corresponds to Fig. \ref{fig6}(c). The
perturbation strength is $A=0.1$. In both cases the breather moves
from the right to the left and approaches the defect from the loss
side. The solid line shows the case when the maxima of the kinetic
energy almost do not catch the lossy region of the defect but one
of the maxima takes place near the maximum of the gain region. As
a result, the breather gains more energy than it loses and it
splits into a kink-antikink pair, so that the kinetic energy does
not oscillate after the breather passes the defect. The dashed
line shows the case where one maximum of the breather's kinetic
energy nearly fits to the maximal loss and the next maximum nearly
fits to the maximal gain. In this case the breather passes through
the defect almost unchanged. Hence, clearly the interplay of the
kinetic energy oscillation with the spatial distribution of the
gain-loss profile is critical in determining the observed
breather-defect interaction phenomenology.

For large perturbations (i.e., stronger defects)
the maxima of the kinetic energy always
catch the lossy region of the defect. Consequently, the breather
do not gain more energy than it losses and it never splits into a
kink-antikink pair for any initial position of the breather. (see
Fig.~\ref{fig6} (d)).

\begin{figure}
\includegraphics[width=10cm]{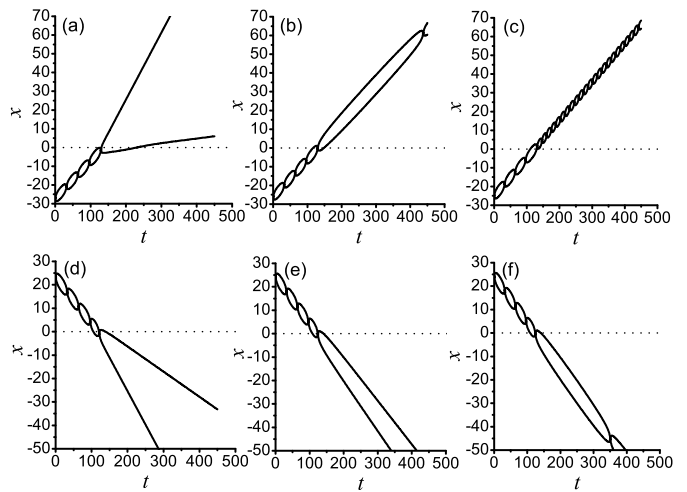}
\caption{Examples of breather dynamics during the interaction with
the defect of strength $A=0.1$. In (a-c) the breather moves toward
the defect from the gain side and in (d-f) from the loss side.
Initial breather parameters are $\omega=0.1$, $V_b=\pm 0.2$ and
initial positions are (a) $x_{0}=-27.34$, (b) $x_{0}=-26.32$, (c)
$x_{0}=-25.16$, (d) $x_{0}=23.48$, (e) $x_{0}=24.40$ and (f)
$x_{0}=24.42$. Horizontal dashed lines show the position of the
defect center and wavy lines represent the breather's two
subkinks.} \label{fig8}
\end{figure}

Examples of the breather interaction with the defect are presented
in Fig.~\ref{fig8} for different initial breather positions. In
(a-c) the breather approached the defect from the gain side and in
(d-f) from the loss side. The breather parameters are $V_b=\pm
0.2$ and $\omega=0.1$ and the perturbation amplitude is $A=0.1$ in
all cases. In (a),(d) and (e) the breather breaks up into
subkinks. The breaking up takes place only for breathers with
sufficiently small frequencies. In (b) and (f) after the
interaction of the defect breather frequency decreases which means
that the breather total energy increases. In (c) the breather
frequency increases (total energy decreases). This again corroborates
the fact that the breather may either lose or gain energy
upon its interaction with the defect (contrary to what we
saw, e.g., in the case of the kink).

\subsection {Kink-antikink-defect interaction} \label{Sec:KinkAntikink}

\begin{figure}
\includegraphics[width=12cm]{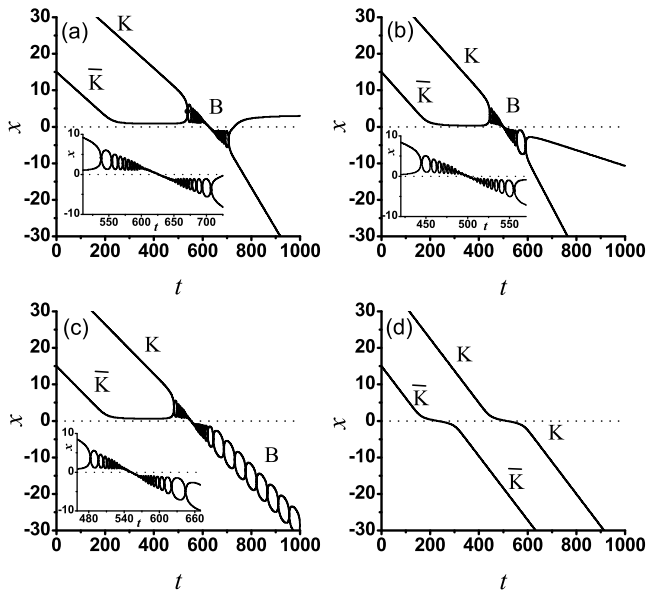}
\caption{Examples of the interaction of a kink-antikink pair with
a defect of strength $A=0.5$.  In all cases, the kink and antikink
move toward the defect from the loss side with equal velocity.
Horizontal dashed lines show the location of the center of the
defect. In (a-c) the velocity of the kinks is smaller than the
threshold value $V_c=-0.0765$: (a) $V_k=-0.06$ (b) $V_k=-0.075$
(c) $V_k=-0.068$, while in (d) $V_k=-0.088$, i.e., both
have velocities above the threshold value and overcome
the defect, effectively without interacting. The kink initial
positions are $x_0=15$ for $\overline{\rm K}$ and $x_0=40$ for
${\rm K}$. The insets show the details of the dynamics close to the
defect center.} \label{fig9}
\end{figure}

In Fig.~\ref{fig9} we present the results obtained for the case
when a well separated antikink and kink move toward the defect with
the velocity $V_k$ from the lossy side. The defect strength is
$A=0.5$ and thus, the kink critical velocity is $V_c=-0.0765$. We
take (a) $V_k=-0.06$ (b) $V_k=-0.075$ (c) $V_k=-0.068$ and (d)
$V_k=-0.088$, so that in the first three cases $V_k<V_c$ and the
antikink is trapped at the lossy region. Then the kink approaches
the antikink and they create a breather. Interestingly, the
breather easily enters the gain side of the defect and it is
amplified. In (a) the breather splits into a kink-antikink pair
with one subkink trapped by the loss region and another one
passing through the defect. In (b) both subkinks pass through the
defect. In (c) the breather does not split and it moves away from
the defect as a single entity (i.e., the antikink and kink
remain bound). In (d) $V_k>V_c$ and the antikink is not trapped by
the defect and both subkinks pass through the defect effectively without
interaction with each other.

\begin{figure}
\includegraphics[width=10cm]{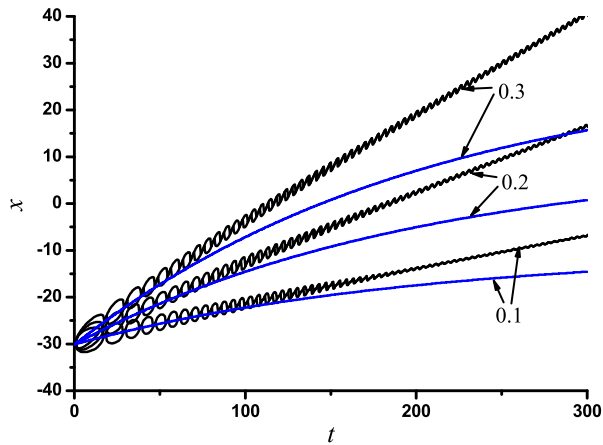}
\caption{Dynamics of breathers and kinks with initial
velocities equal to 0.1, 0.2 and 0.3 in the case of homogeneous
loss $\gamma(x)\equiv 1$ and $A=-0.005$. The breather initial
frequency is $\omega=0.1$. The kink trajectories (smooth lines)
show that kink propagation velocity gradually decreases, while
breather trajectories (wavy lines) demonstrate almost constant
propagation velocities, i.e., minimal impact in the breather
translation by the presence
of the dissipative perturbation.} \label{fig10}
\end{figure}

Note that in Fig.~\ref{fig9}(b) both kink and antikink have
$V_k<V_c$ nevertheless, they {\em both} pass through the defect.
Two reasons can be given to explain this effect and they both are
related to the fact that the kink and antikink form a breather to
pass through the defect. In Sec. \ref{Sec:Breather} it was shown
that the breather can gain energy from the
$\mathcal{PT}$-symmetric defect, depending on the phase, and this
is the first reason. The second reason is that the breather
translational degree of freedom is only weakly affected by the
perturbation considered in this study. This is demonstrated in
Fig.~\ref{fig10} where we contrast the dynamics of breathers and
kinks with initial velocities equal to 0.1, 0.2 and 0.3 in the
case of homogeneous loss $\gamma(x)\equiv 1$ and $A=-0.005$. The
breather initial frequency is $\omega=0.1$. It can be seen that
the kink trajectories (smooth lines) show that kink propagation
velocity gradually decreases, while breather trajectories (wavy
lines) demonstrate almost constant propagation velocities. Hence,
this suggests that while breathers are topologically robust,
breathers are more efficient in weathering lossy media and in
overcoming barriers imposed by dissipative perturbations. This is
a feature that is especially useful in the realm of
$\mathcal{PT}$-symmetric perturbations/defects.

\section {Conclusions} \label{Sec:V}

Interaction of SG kinks (and multi-kinks) as well
as breathers with a
$\mathcal{PT}$-symmetric defect bearing balanced regions of
positive and negative dissipation of energy was investigated
analytically (wherever possible) and numerically in the present work.

It was demonstrated that a kink coming from the gain side
always passes through the defect and restores its initial velocity
(see Fig.~\ref{fig1}). The only effect of the interaction with the
defect is a phase shift associated with
the kink position. However, for the kink
approaching the defect from the opposite side, there exist two
different scenarios, depending on the kink initial velocity $V_k$.
For $V_k<V_c$, where $V_c$ is a threshold value of the velocity,
the kink does not have enough energy to pass through the defect
and it is trapped by the lossy side of the defect (see
Fig.~\ref{fig2}), while for $V_k>V_c$ it is able to overcome
the relevant barrier.

If two well-separated kinks approach the defect from the lossy
side with the velocities less than $V_c$, then one of them can
pass through the defect while another one will be either trapped
by the lossy region or reflected back (see Fig.~\ref{fig5}), i.e.,
their pairwise repulsion may modify the collisional outcome
with the defect.

The breather-defect interaction is more interesting since the
breather can split into subkinks depending on its parameters and
also on the amplitude of the defect. Depending on the breather
initial phase, its total energy can be increased or decreased
after the interaction with the defect (see Fig.~\ref{fig6} and
Fig.~\ref{fig8}). This can be explained by the fact that the
kinetic energy of the moving breather is a periodic function of
time and space with sharp maxima at the points where the
subkinks collide. The type of perturbation considered in the
present work is more prominent for large $\phi_t$. Change in the
breather phase changes the location of the subkink collision
points with respect to the maxima of the gain and loss regions
thus affecting the overall result of the interaction between the
breather and the defect (see Fig.~\ref{fig7}).

A well-separated kink and antikink pair moving toward the lossy side of
the defect with $V_k<V_c$, may enable {\em both} coherent structures
to potentially pass through the defect
[see Fig.~\ref{fig9}(b)]. This happens because the kink and
antikink form a breather that can gain energy from the defect and
whose propagation velocity is less affected by the dissipative
term than the propagation velocity of the constituent kink or antikink
(see Fig.~\ref{fig10}).

We conclude that the $\mathcal{PT}$-symmetric defects give new
opportunities in the manipulation with the soliton dynamics in the
sine-Gordon equation and related field theories. Numerous future directions
open up as a result of the present considerations.
One such is to consider other Klein-Gordon field theories
in the presence of $\mathcal{PT}$-symmetric defects, such
as, e.g., the $\phi^4$ model. The latter is especially interesting
due to the presence of internal modes in the kink dynamics
which may have a nontrivial impact on the observed phenomenology.
Another relevant consideration is that of higher dimensionality.
Examining radial kinks as well as breathers in the higher
dimensional versions of the sine-Gordon model is a theme
that has attracted recent interest~\cite{jeanguy}, including
the formation of breathers as a result of the interaction of
the kinks with a radial domain boundary. Developing $\mathcal{PT}$-symmetric
variants of the 2d sine-Gordon and examining the corresponding
dynamics is still an open problem. Finally, comparison of the
present features with corresponding bright and dark soliton
interactions with $\mathcal{PT}$-symmetric defects within the
realm of the focusing and defocusing respectively $\mathcal{PT}$-symmetric
nonlinear Schr{\"o}dinger equation would also be a theme of
relevance to future studies, especially since the latter
is the principal field of optical applications of $\mathcal{PT}$-symmetric
models. Such studies are presently under consideration and will
be reported in future publications.

\section*{Acknowledgments}

D.S. thanks the hospitality of the Bashkir State Pedagogical
University and the Institute for Metals Superplasticity Problems,
Ufa, Russia. S.V.D. thanks financial support provided by the
Government Program 5-100-2020 and by the Russian Science
Foundation grant 14-13-00982. D.I.B. was partially supported by a
grant of Russian Foundation for Basic Research, the grant of the
President of Russian Federation for young scientists-doctors of
science (project no. MD-183.2014.1) and by the fellowship of
Dynasty foundation for young Russian mathematicians.
P.G.K. acknowledges support from the US National Science
Foundation under grants CMMI-1000337, DMS-1312856, from FP7-People under
grant IRSES-606096 from
the Binational (US-Israel) Science Foundation through grant 2010239, and from
the US-AFOSR under grant FA9550-12-10332.

\end{document}